\documentclass{aa}
\usepackage{graphicx}
\usepackage{longtable}
\usepackage{latexsym}
\usepackage{amssymb}
\usepackage{lscape}
\usepackage{natbib}
\begin{document}
   \title{Multiple merging in the Abell cluster 1367.\thanks{Based on observations obtained with the 
   William Herschel Telescope operated on the island of La Palma (Spain) by the Isaac Newton Group, with 
   the Loiano telescope belonging to the University of Bologna (Italy) and with the G.Haro telescope of the INAOE (Mexico).}}

   \author{L.Cortese\inst{1}, G.Gavazzi\inst{1}, A.Boselli\inst{2}, J.Iglesias-Paramo\inst{2},\and L.Carrasco\inst{3,4}}

   \offprints{L.Cortese}

   \institute{Universit\'{a} degli Studi di Milano-Bicocca, P.zza della Scienza 3, 20126 Milano, Italy.\\
              \email{Luca.Cortese@mib.infn.it; Giuseppe.Gavazzi@mib.infn.it}
         \and
             Laboratoire d'Astrophysique de Marseille, BP8, Traverse du Siphon, F-13376 Marseille, France.\\
             \email{alessandro.boselli@oamp.fr; jorge.iglesias@oamp.fr}
          \and
	     Instituto Nactional de Astrofisica, Optica y Electronica, Apartado Postal 51 C.P. 72000 Puebla, Pue., Mexico.
	     \email{carrasco@transum.inaoep.mx}
	  \and
	     Observatorio Astronomico Nacional/UNAM, Ensenada B.C., Mexico.	           
             }

   \date{Received 4 March 2004 | Accepted 4 June 2004}

\abstract{We present a dynamical analysis of the central $\sim$ 1.3 square degrees of the 
cluster of galaxies Abell 1367, based on 273 redshift measurements (of which 119 are news). 
From the analysis of the 146 
confirmed cluster members we derive a significantly non-Gaussian velocity distribution, with 
a mean location $C_{BI} = 6484\pm81 ~\rm km~s^{-1}$ and a scale $S_{BI} = 891\pm58~\rm km~s^{-1}$. 
The cluster appears elongated 
from the North-West to the South-East with two main density peaks associated 
with two substructures. 
The North-West subcluster is probably in the early 
phase of merging into the South-East substructure ($\sim$ 0.2 Gyr before core crossing).\\
A dynamical study of the two subclouds points out the existence 
of a group of star-forming galaxies infalling into the core of the South-East subcloud 
and suggests that two other groups are infalling into the NW and SE 
subclusters respectively.
These three subgroups contain a higher fraction of star-forming galaxies
than the cluster core, as expected during merging events.
Abell 1367 appears as a young cluster currently forming at the 
intersection of two filaments.

}

\titlerunning{Multiple merging in the Abell cluster 1367}
\authorrunning{L.Cortese et al.}

   \maketitle
%

\section{Introduction}

Clusters of galaxies represent the most massive gravitationally 
bound systems in the Universe. 
They provide us with valuable insights into the formation 
of large-scale structures, as well as into
the formation and evolution of galaxies.
The hierarchical model predicts 
that galaxy clusters are formed by accretion of units 
of smaller mass at the nodes of large-scale filaments \citep{WESVI91,KATWH93}.
Statistical analyses of clusters have shown 
that even at low redshift a high fraction of clusters presents substructures, 
implying that clusters are still dynamically 
young units, undergoing the process of formation \citep{DRESS88}.\\
The Abell cluster 1367 ($z\sim0.0216$) 
lies at the intersection of two filaments, the first extending 
roughly 100 Mpc from Abell 1367 toward Virgo \citep{WEST00}, the second connecting 
Abell 1367 to Coma (as a part of the Great Wall, \citealp{ZAGE93}). 
With its irregular X-ray distribution \citep{JONES79,BE83,GREB95}, 
high fraction of spiral galaxies and low central galaxy density, Abell 1367 can be 
considered as the prototype of a nearby dynamically young cluster.\\
ASCA X-ray observations pointed out the existence of a strong localized shock 
in the intra-cluster medium (ICM) suggesting that Abell 1367 is experiencing a merging between 
two substructures \citep{DON98}.
Moreover recent  Chandra observations \citep{SUNM02}, and a preliminary 
analysis of the XMM data \citep{FORCH03},  indicate the presence of 
cool gas streaming into the cluster core, supporting a multiple merger scenario.\\
Optical and radio observations also suggest that this cluster is currently 
experiencing galaxy infall into its center. 
\cite{GAVCO95,GAVB01} discovered two head-tail radio sources associated with disk galaxies with an excess 
of giant HII regions on their leading edges, in the direction of the cluster center.
The observational scenario is consistent with the idea that ram-pressure 
\citep{GUNG72} is, for a limited amount of time, 
enhancing the star formation of galaxies that are entering the 
cluster medium.
In addition \cite{GAVC03} pointed out the existence of a group of star bursting galaxies 
infalling into the cluster core.\\
Although X-ray, radio and optical observations suggest that Abell 1367 is 
dynamically young and it is still undergoing the process of formation, 
detailed spatial and dynamical analysis of this cluster has not been attempted so far.
\cite{GIRARD98} detected a secondary peak in the cluster velocity 
distribution, suggesting that Abell 1367 is a binary cluster, but their analysis
was based on $\sim$ 90 redshifts, insufficient for drawing a detailed model of the cluster kinematics.\\
\cite{CORGAV03} carried out a deep ($r'<20.5$)
spectroscopic survey of the central $\sim$ 1.3 square degrees of Abell 1367
adding 60 new spectra (33 members). 
Here we present new measurements for 119 galaxies (adding another 33 cluster members). 
In total 273 redshifts were measured in the region, out of which 146 are cluster members,
allowing the first detailed dynamical analysis of Abell 1367.\\
This paper is arranged as follows: Sec.\ref{sample} briefly describes the studied sample and 
it contains a description of the observations 
and data reduction. The analysis of the 1D velocity distribution is given in Sec.\ref{1d}. 
Tests of 3D substructure are presented in Sec.\ref{3d}. 
The various spatial and/or velocity substructures detected in our 
sample are studied in Sec.\ref{galdistr}. The star formation of galaxies in the infalling groups 
is discussed in Sec.\ref{star}. 
The mass of the whole cluster and of the two main substructures are computed in Sec.\ref{mass}.
Sec.\ref{2b} studies the current dynamical state of the system.
Our conclusions are briefly summarized in Sec.\ref{concl}.\\
We assume a cluster distance of 91.3 Mpc \citep{GAVCAR99} corresponding to a Hubble 
constant of 71 $\rm km~s^{-1}~Mpc^{-1}$. 

\section{Observations and data reduction}
\label{sample}
\begin{figure}[t]
\centering
\includegraphics[width=8.5cm]{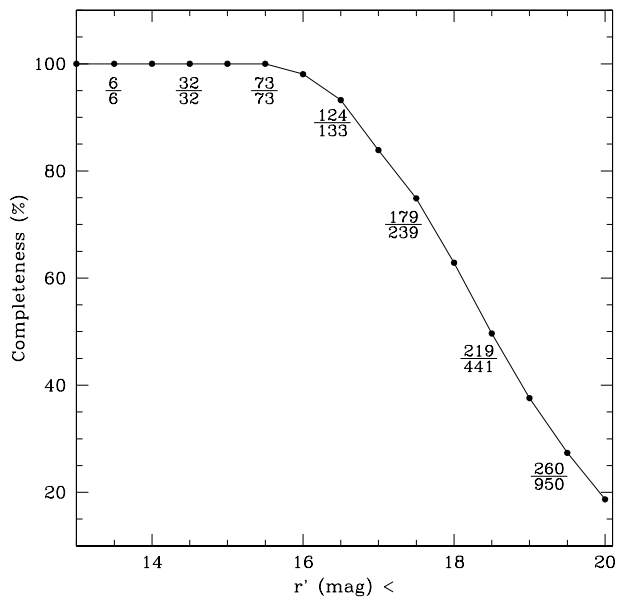}
\caption{Cumulative redshift distribution for galaxies in the studied region.}
\label{completeness}
\end{figure}

The cluster region analyzed in this work covers an area of $\sim$ 1.3 square degrees 
centered at $\alpha(J.2000)= \rm 11h44m00s$ $\delta(J.2000)= \rm 19d43m30s$. 
$r'$ imaging material was used to extract a catalogue of galaxy candidates  
in Abell 1367 complete to $r'\sim20.5$ mag, and to select the targets of the present spectroscopic survey.
The majority of the photometric catalogue, covering the northern and western part of Abell 1367, 
was published by \cite{IPB03}; conversely 
the south-east catalogue extension will be given in a forthcoming paper.\\
Spectroscopy of Abell 1367 was obtained with the 
AF2-WYFFOS multi fiber spectrograph at the 4.2m William 
Herschel Telescope (WHT) on La Palma (Spain) during 2003, March 27-29. 
WYFFOS has 150 science fibers of 1.6 arcsec diameter 
coupled to a bench-mounted spectrograph which relies on a TEK $\rm 1024\times1024$ CCD. 
The 316R grating was used, giving a dispersion of 
$\sim$240 $\rm \AA / mm$, a resolution of $\sim6 \rm \AA$ FWHM, and a total spectral coverage 
of $\sim$5600 $\rm \AA$. The spectra were 
centered at $\sim6500 \rm \AA$, thus covering from $\rm 3600~ \AA~ to~ 9400~ \AA$. 
We allocated typically $\sim$ 70 objects to fibers 
in a given configuration and, on average, 15 sky fibers. 
A total of 4 configurations were executed, with an 
exposure time of 4x1800 sec for each configuration. 
Argon lamps for wavelength calibration were obtained for each exposure.\\ 
The reduction of the multi fiber spectra was performed in the IRAF\footnote{IRAF is 
distributed by the National Optical Astronomy Observatories, which is operated 
by the Association of Universities for Research in Astronomy, Inc., under 
the cooperative agreement with the National Science Foundation} environment, 
using the IMRED package.
After bias subtraction, the apertures were defined on dome flat-field frames and used to 
trace the spectra on the CCD. 
The arc spectra were extracted and matched with arc lines to determine the dispersion solution. 
The rms uncertainty of the wavelength calibration ranged between 0.1 and 0.3 $\rm\AA$.
The lamps' wavelength calibration was checked 
against known sky lines. These were found within $\rm \sim 0.5~\AA$ of their nominal 
position, providing an estimate of the systematic
uncertainty on the derived velocity of $\rm \sim 25~km~s^{-1}$.  
The object spectra were extracted, wavelength calibrated and normalized
to their intensity in the interval 5400-5600 $\rm \AA$.
A master sky spectrum, that was constructed by combining various
sky spectra was normalized to each individual science spectrum and then
subtracted from it. Unfortunately strong sky residuals were left after this procedure, limiting the
number of useful spectra to 98 (as listed in Tab. \ref{redshift}).\\
Nine additional long-slit, low dispersion spectra were obtained in March 2003 and in February 2004 using the 
imaging spectrograph BFOSC attached to the Cassini 1.5m telescope at Loiano (Italy). 
Another twelve spectra were taken with LFOSC at the 2.1m telescope of the Guillermo
Haro Observatory at Cananea (Mexico). 
These observations were performed using a 2.0 arcsec slit and 
the wavelength calibration was secured with exposures of HeAr and XeNe lamps at Loiano 
and Cananea respectively. 
The on-target exposure time ranged between 15 and 30 min according to the brightness 
of the targets.
After bias subtraction, 
when 3 or more frames of the same target 
were obtained, these were combined (after spatial alignment) using a median filter 
to help cosmic rays removal. 
Otherwise the cosmic rays were removed using the task COSMICRAYS and/or under 
visual inspection. 
The lamps wavelength calibration was checked 
against known sky lines. These were found within $\rm \sim 1~\AA$ from their nominal 
position, providing an estimate of the systematic
uncertainty on the derived velocity of $\rm \sim 50~km~s^{-1}$. 
After subtraction of sky background, one-dimensional spectra were 
extracted from the frames.\\ 
The redshift were obtained using the IRAF FXCOR Fourier 
cross-correlation \citep{TODA79} 
task, excluding the regions of the spectra affected 
by night-sky lines. 
Moreover all the spectra and their best correlation function were visually examined  
to check the redshift determination.\\
Table \ref{Tab1} lists the characteristics of the instrumentation 
in the adopted set-up. \\
The 119 new velocity measurements presented in this work are listed in Table \ref{redshift} 
as follows:\\
Column 1: Galaxy designation.\\
Column 2, 3: (J2000) celestial coordinates, measured with few arcsec uncertainty.\\
Column 4: $r'$ band magnitude.\\
Column 5: observed recessional velocity.\\
Column 6: telescope (WHT=William Herschel Telescope; LOI=Loiano; CAN=Cananea)\\
Combining the new set of 119 redshifts (given in Tab. \ref{redshift}) with 
the ones available from the literature (NED; \citealp{CORGAV03, RINGE03}), we have the redshift for 273 galaxies
of which 146 are cluster members 
($\rm 4000~ km~ s^{-1} \leq \it V \rm \leq 10000~ km~ s^{-1}$).\\
The cumulative redshift distribution, in the observed
area, as a function of the $r'$ magnitude is shown in Fig.\ref{completeness}.
The completeness is $\sim$ 70\% at $r'<17.5$, and it drops 
to $\sim$ 45\% at $r'<18.5$.

\begin{table*}  
\caption{The spectrograph characteristics} 
\label{Tab1} \[ 
\begin{tabular}{lccccccc} 
\hline 
\noalign{\smallskip}   
Observatory &  Dates & N. gal.& Spectrograph & Dispersion &  Coverage  & CCD & pix   \\      
	&  & & &$\rm \AA/mm$ & $\rm \AA$ &  & $\rm \mu m$ \\ 
\noalign{\smallskip} 
\hline 
\noalign{\smallskip} 
WHT       & March~03 & 98 & AF2-WYFFOS & 240 & 3600-9400  & $1024\times1024~TEK$ & 24 \\
Cananea   & March~03 & 12 & LFOSC      & 228 & 4000-7100  & $576\times384~TH$    & 23 \\   
Loiano    & March~03 - Feb.~04 & 9 & BFOSC      & 198 & 3600-8900  & $1340\times1300~EEV$ & 20 \\
\noalign{\smallskip} 
\hline 
\end{tabular} \] 
\end{table*}

\section{The global velocity distribution}
\label{1d}
\begin{figure}[t]
\centering
\includegraphics[width=8.5cm]{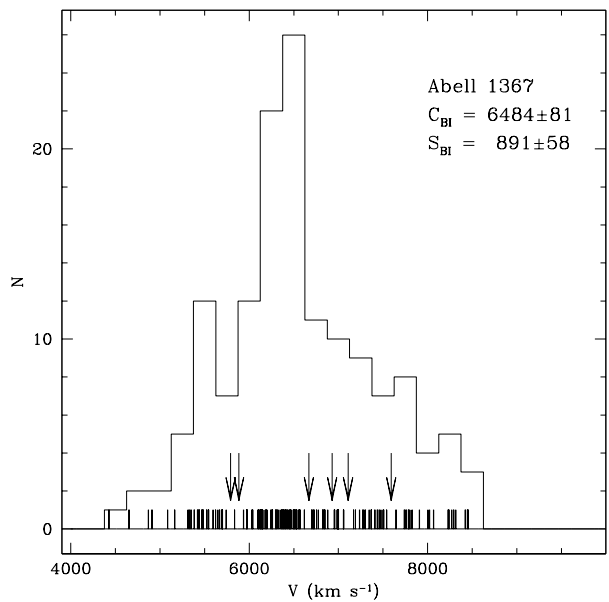}
\caption{Velocity histogram and stripe density plot for the members of Abell 1367. Arrows 
mark the location of the most significant weighted gaps in the velocity distribution.}
\label{tothist}
\end{figure}
The line of sight (LOS) velocity distribution for the 146 cluster members 
is shown in Fig. \ref{tothist}.
The mean and standard deviation are known to be efficient 
estimators of the central location and scale when the underlying 
population is gaussian. Unfortunately 
they are not minimum variance estimators when the nature 
of the observed population is significantly non-Gaussian.
The best location and scale estimators must be resistant 
to the presence of outliers and robust to a broad range 
of non-Gaussian underlying populations.
Thus, following \cite{BEERF90}, we consider the biweight estimator as 
the best estimator of location ($C_{BI}$) and scale ($S_{BI}$) of the cluster velocity distribution.\\ 
We find a location $C_{BI} = 6484\pm81 ~\rm km~s^{-1}$ and a scale $S_{BI} = 891\pm58~\rm km~s^{-1}$,
in agreement with previous studies (e.g. \citealp{GIRARD98,STROOD99}).
Visual inspection of Fig. \ref{tothist} suggests that
the velocity distribution differs from a Gaussian, a deviation that should be quantified 
using appropriate statistical tests.\\
We analyze the higher moments of the distributions using the kurtosis and 
the skewness shape estimators.
Kurtosis indicates a difference in the tails length compared to a Gaussian
(positive kurtosis is indicative of long tails). 
Skewness indicates asymmetry
(positive skewness implies that the distribution is depleted from values lower than the mean location, 
conversely negative skewness denotes a depletion of values higher than the mean).\\ 
In addition we calculate the asymmetry index (AI) and tail index (TI) 
introduced by \cite{BIBE93} as alternatives to the distribution higher moments.
These indicators measure the shape of a distribution but, contrary to  
skewness and kurtosis, which depend on the estimate of the location and the scale 
of the underlying distribution, they are based on the order statistics of the dataset.
The AI measures the symmetry in a population by comparing gaps in the data on the left 
and right sides of the sample median. The TI compares the spread of the dataset 
at 90\% level to the spread at the 75\% level.\\ 
The kurtosis, skewness and the TI reject a Gaussian distribution
with a confidence level of $\geq$99\%, suggesting that the cluster velocity distribution has longer tails  
than a Gaussian of the same dispersion.
Moreover, in order to assess the normality of the velocity distribution, we 
use the Wilk - Shapiro (W) test \citep{YAHIL77}.
Contrary to the $\chi^2$ and Kolmogorov Smirnov, this 
test does not require any hypothesis on the mean and variance of the normal distribution.
The W test rejects normality with a confidence level of 98.7\%, in agreement 
with kurtosis, skewness and TI (see Table \ref{1dstat}).\\ 
The departure from a normal distribution could result 
from a mixture of several velocity distributions with different location 
and smaller velocity dispersion than the whole sample; 
thus, using 
the program ROSTAT \citep{BEERF90}, we investigate the presence of significant 
gaps \citep{BEGE91} in the velocity 
distribution, indicating subclustering.
A weighted gap is defined by:
\begin{displaymath}
y_{i} = \Big(i(N-i) * (x_{i+1}-x_{i})\Big)^{1/2}
\end{displaymath}
where $N$ is the number of values in the dataset.
A weighted gap is significant if its value, 
relative to the midmean (the mean of the central 50\% of the dataset) 
of the other weighted gaps, is greater than 
2.25. This value corresponds to a probability of occurrence in a normal 
distribution of less than 3\%.
We detected six significant weighted gaps in the Abell 1367 velocity distribution.
The stripe density plot of radial velocities and the position of each gap 
(indicated with an arrow) are shown in Fig. \ref{tothist}.
The velocity of 
the object preceding each gap, the normalized size of the gap and 
the probability of finding 
a normalized gap of the same size and position in a normal distribution 
are listed in Table \ref{gap}.
\begin{table}  
\caption{1D substructure indicators for the whole cluster sample} 
\label{1dstat} \[ 
\begin{tabular}{ccc} 
\hline 
\noalign{\smallskip}   

Test   	  & Value    & Rejection of a gaussian \\
\noalign{\smallskip} 
\hline 
\noalign{\smallskip}
AI   	   & -0.077  & $\leq80$ \% \\
TI         & 1.240  & $>$99 \% \\
Skewness   & 0.269  & $>$99 \%   \\
Kurtosis   & 2.680  & $>$99 \%   \\	
W    	   & 0.963  &  98.7 \% \\
\noalign{\smallskip} 
\hline 
\end{tabular} \] 
\end{table}

\begin{table}  
\caption{The most significant weighted gaps detected in the velocity distribution 
of the whole cluster sample.} 
\label{gap} \[ 
\begin{tabular}{ccc} 
\hline 
\noalign{\smallskip}   
Velocity & ~Gap~ & Significance \\      
$\rm ~km~s^{-1}$  &        &  \\ 
\noalign{\smallskip} 
\hline 
\noalign{\smallskip} 
5742  &    2.53  &  1.40\% \\
5835  &    2.66  &  1.40\% \\
6619  &    2.90  &  0.60\% \\
6880  &    2.64  &  1.40\% \\
7059  &    3.01  &  0.20\% \\
7542  &    2.33  &  3.00\% \\
\noalign{\smallskip} 
\hline 
\end{tabular} \] 
\end{table}  

\section{Localized velocity structures}
\label{3d}
Given the non-Gaussian nature of the velocity distribution, 
we looked for spatially localized variations 
in the LOS velocity and velocity dispersion distributions. 
First of all we applied the three 3D tests 
commonly used to quantify the amount of substructures in galaxy clusters:
the $\rm \Delta$ test \citep{DRESS88}, the $\alpha$ test \citep{WESB90} and the $\epsilon$ test 
\citep{BIRD94}.\\
The $\Delta$ test is based on the comparison of the local mean velocity, $V_{local}$, 
and the velocity dispersion, $\sigma_{local}$, associated to each cluster member 
(computed using its 10 nearest neighbors) with 
the mean velocity $V$, and dispersion $\sigma$, of the whole galaxy sample.
For each galaxy, the deviation is defined by:
\begin{displaymath}
\delta^{2}= \frac{11}{\sigma^{2}}[(V_{local} - V)^{2} + (\sigma_{local}-\sigma)^{2}]
\end{displaymath}
The observed cumulative deviation $\Delta$, defined as the sum of the $\delta$'s for 
the cluster members, is used to quantify the presence of substructures.
As shown by \cite{PINK96} for samples with no substructures, the value of $\Delta$ is 
approximately equal to the total number of galaxies,  while it is larger in the presence of substructures.\\
The $\alpha$ test measures how much the centroid of the galaxy distribution 
shifts as a result of correlations between the local kinematics and the projected 
galaxy distribution.
The centroid of the whole galaxy distribution is defined as:
\begin{displaymath}
x_{c} = \frac{1}{N} \sum_{i=1}^{N} x_{i} ~~~~~~~~~  y_{c} = \frac{1}{N} \sum_{i=1}^{N} y_{i}  
\end{displaymath}
For each galaxy $i$ and its 10 nearest neighbors in the velocity space, 
the spatial centroid is defined as:
\begin{displaymath}
x_{c}^{i} = \frac{\sum_{j=1}^{11} x_{j}/\sigma_{j}}{\sum_{j=1}^{11} 1/\sigma_{j}} ~~~~~~~~~
y_{c}^{i} = \frac{\sum_{j=1}^{11} y_{j}/\sigma_{j}}{\sum_{j=1}^{11} 1/\sigma_{j}} 
\end{displaymath}
where $\sigma_{j}$ is the velocity dispersion for galaxy $j$ and its 10 
nearest neighbors in projection.
Finally the presence of substructures in the cluster sample is quantified using 
the $\alpha$ statistic defined as: 
\begin{displaymath}
\alpha = \frac{1}{N} \sum_{i=1}^{N} [(x_{c}^{i} - x_{c})^2 +  (y_{c}^{i} - y_{c})^2]^{1/2}
\end{displaymath}
which represents the mean centroid shift for the galaxy cluster.
The higher the value of $\alpha$, the higher the probability of 
substructures.\\
The $\epsilon$ test quantifies the correlations between the position and the projected mass 
estimator \citep{HETRE85}, defined as:
\begin{displaymath}
M_{PME} = \Big(\frac{32}{\pi G N}\Big) \sum_{j=1}^{N} v_{zj}^{2} r_{j}
\end{displaymath} 
where $v_{zj}$ is the radial peculiar velocity with respect to the nearest neighbors 
group (composed by a galaxy and its 10 nearest neighbors) and $r_{j}$
is the projected distance from the center of the nearest neighbor 
group. The substructure statistic is then defined as:
\begin{displaymath}
\epsilon = \frac{1}{N_{gal}} \sum_{i=1}^{N} M_{PME}
\end{displaymath}
which represents the average mass of the nearest neighbors groups in the cluster.
Since galaxies in the nearest neighbors 
groups have small projected separations, $\epsilon$ is generally 
smaller than the global mass estimate.
$\epsilon$ is lower for a cluster with substructures 
than for a relaxed system.\\ 
The value and the significance 
of the above tests are listed in Table \ref{3Dsub}.
\begin{table}  
\caption{3D substructure indicators for our sample} 
\label{3Dsub} \[ 
\begin{tabular}{ccc} 
\hline 
\noalign{\smallskip}   
Indicator & Value & Prob. of substructures \\      
\noalign{\smallskip} 
\hline 
\noalign{\smallskip} 
$\rm \Delta$    & 206.5     &  99.8 \%	       \\
$\rm \alpha$    & 0.161~Mpc & 55.7 \%           \\
$\rm \epsilon$  & $5.44~10^{13}~\rm M_{\odot}$  & 68.4 \% 	       \\
\noalign{\smallskip} 
\hline 
\end{tabular} \] 
\end{table}  
These statistical tests 
are calibrated using 1000 Monte Carlo simulations that randomly shuffle 
the velocity of galaxies, keeping fixed their observed position, thereby destroying
any existing correlation between velocity and position.
The probability of subclustering is then given as the fraction 
of simulated clusters for which the test value is lower (larger for the $\epsilon$ test) than 
the observed one.   
\begin{figure}[t]
\centering
\includegraphics[width=9.0cm]{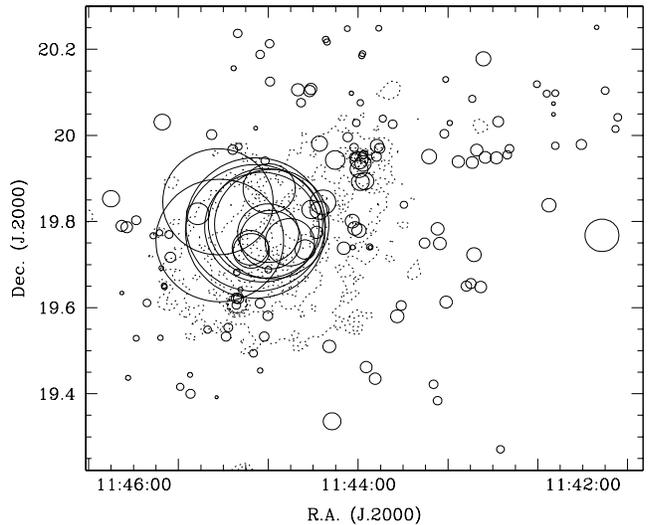}
\caption{Local deviations from the global kinematics for galaxies in Abell 1367 as measured 
by the \cite{DRESS88} test. Galaxies are marked with 
open circles whose radius scales with their local deviation $\delta$ from the global kinematics.
The ROSAT X-ray contours are shown with dotted lines.}
\label{dressler}
\end{figure}
Assuming that these tests reject the null hypothesis if the confidence level 
is greater than 90\%, only the $\Delta$ test finds evidence of substructures (see Table 
\ref{3Dsub}). The local deviations from 
the global kinematics as measured by the $\Delta$ test are shown in Fig \ref{dressler}. 
The positions of galaxies are marked with 
open circles whose radius scales with their 
local deviation $\delta$ from the global kinematics.
The presence of a substructure with a high deviation from 
the global cluster kinematic is evident projected near the cluster core.\\
More insights on the cluster dynamical state can be achieved by comparing the results 
of the one and three dimensional statistical tests with the 
N-body simulations performed by \cite{PINK96}. 
These authors analyzed how the significance level 
of statistical tests of substructure varies in different cluster merging scenarios.
The deviation of the velocity distribution from a Gaussian and the detection of substructure  
provided by the $\Delta$ test suggest that Abell 1367 is in the early merging stage, 
$\sim 0.2$ Gyr before core crossing.

\section{The cluster dynamics}
\label{galdistr}
\begin{figure*}[t]
\centering
\includegraphics[width=8.7cm]{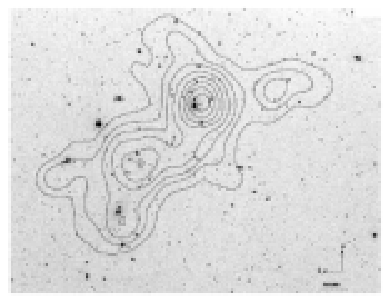}
\includegraphics[width=8.7cm]{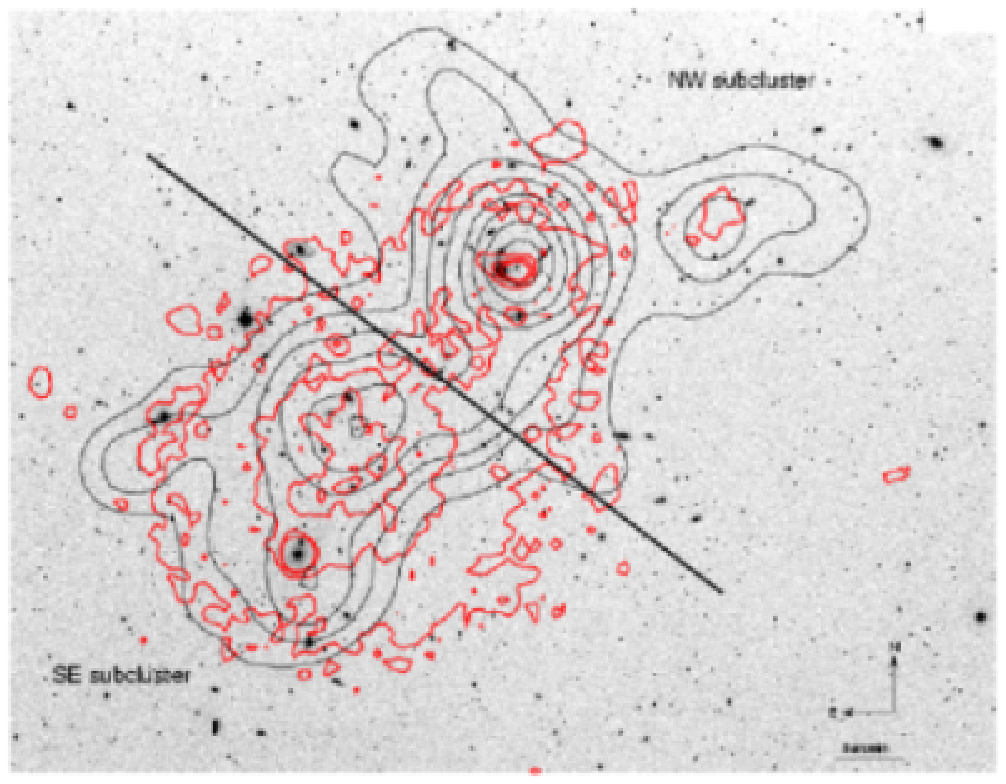}
\caption{Palomar DSS image of the central region ($\sim$1.3 square degrees) of Abell 1367 studied 
in this work. The iso-density contours for the 146 confirmed 
cluster members are superposed. The lowest iso-density contour correspond to 3$\sigma$ above the mean 
density in the field (left).
The ROSAT X-ray contours are superposed in red (right). The straight line 
indicates the position of the abrupt gas temperature gradient detected by ASCA \citep{DON98}, 
used to divide our sample into two subclusters: the North-West and the South-East.}
\label{isodensity}
\end{figure*}

The analysis of the galaxy distribution, of the local mean LOS velocity and of the velocity dispersion
give further insight onto the cluster structure. 
The iso-density map of the cluster members (computed using the 10 nearest neighbors to each point)
is shown in Fig.\ref{isodensity} (left).
The galaxy distribution appears elongated from north-west to south-east 
with two major density peaks.
The highest density region corresponds approximately to 
the center of the NW X-ray substructure detected by 
ROSAT \citep{DON98}, while the 
secondary density peak is slightly offset from the X-ray cluster center 
($\alpha(\rm J.2000)=
11h44.8m ~\delta(\rm J.2000)=19d42m$, \citealp{DON98}).
Moreover the south galaxy density peak roughly coincides with the substructure 
detected by the $\Delta$ test (see Fig.\ref{dressler}) and with the infalling group of 
star-forming galaxies studied by \cite{GAVC03}.\\
\begin{figure*}[t]
\centering
\includegraphics[width=7.5cm]{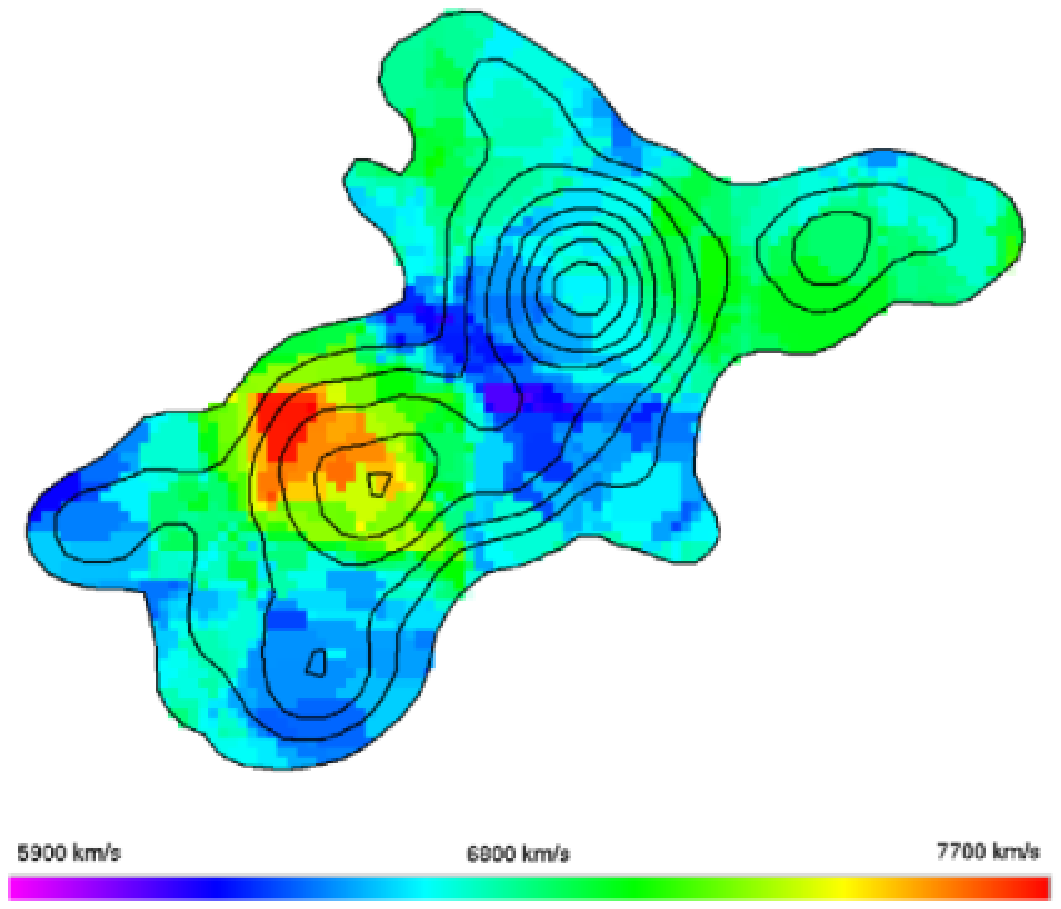}
\includegraphics[width=7.5cm]{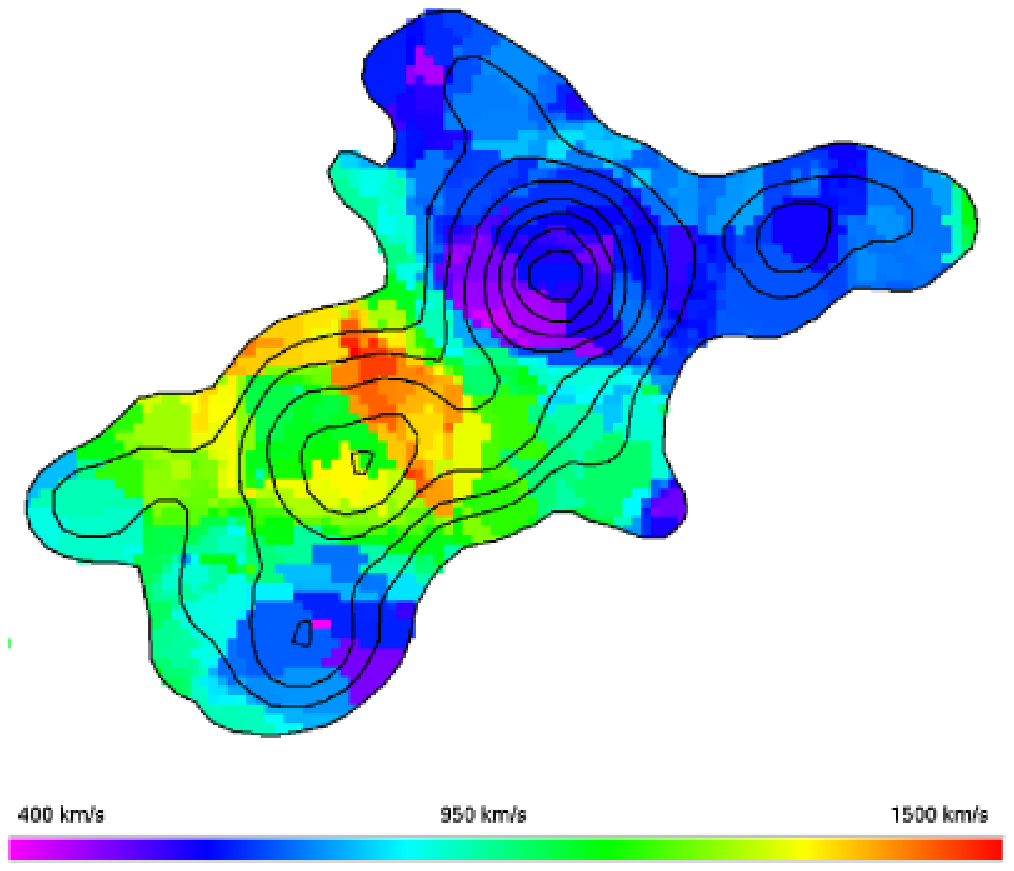}
\caption{The LOS velocity field (left) and the velocity dispersion field (right) for the whole 
region studied in this work. The LOS velocity and the velocity dispersion  
are computed using the 10 nearest neighbors to each pixel, whose size is 36 $\rm arcsec^{2}$.
The iso-density contours for the 146 confirmed cluster members are superposed in black. }
\label{velfield}
\end{figure*}
The iso-density contours superposed on 
the ROSAT X-ray contours are shown in Fig.\ref{isodensity} (right).
The region between the two major density peaks coincides with 
the strong gradient in the gas temperature (see the straight line in Fig.\ref{isodensity}, right) 
observed for the first time by 
ASCA \citep{DON98} and recently confirmed by Chandra \citep{SUNM02}.
This abrupt temperature change is strongly suggestive 
of a shock which has generated during a collision 
between two substructures, probably associated with the SE and the NW galaxy density peaks.
In fact N-body simulations show that temperature structures and X-ray morphology 
similar to the one observed in Abell 1367 are typical of clusters at 
an early merging phase ($\sim 0.25$ Gyr before core crossing) 
(\citealp{SCHIND93,GOMLO02}).\\
The merging scenario is further supported by the LOS velocity 
and velocity dispersion fields (computed using the 10 nearest neighbors to 
each point) shown in Fig. \ref{velfield}.
The SE subcluster has higher LOS velocity and velocity dispersion 
than the NW substructure.
The region with the highest LOS velocity and velocity dispersion lies 
$\sim$ 6 arcmin N from the X-ray cluster center and it coincides with 
the substructure detected by the $\Delta$ test.
This result points out the presence of a group of galaxies infalling 
in the SE cluster core (see Sec.\ref{SE}).\\
Thus the NW subcluster appears as a relaxed system with the lowest velocity dispersion 
among the whole sample; on the other hand the SE subcluster appears far from 
relaxation, and it is probably experiencing a multiple merging event.\\
We use the position of the gas temperature gradient, shown by 
the straight line in Fig.\ref{isodensity} (right), to divide our sample into two regions 
and to study separately the dynamical properties of the two subclusters.\\
A sketch of the cluster dynamical model discussed in the next section 
is given in Fig.\ref{cartoon}.
\begin{figure*}[t]
\centering
\includegraphics[width=14.5cm]{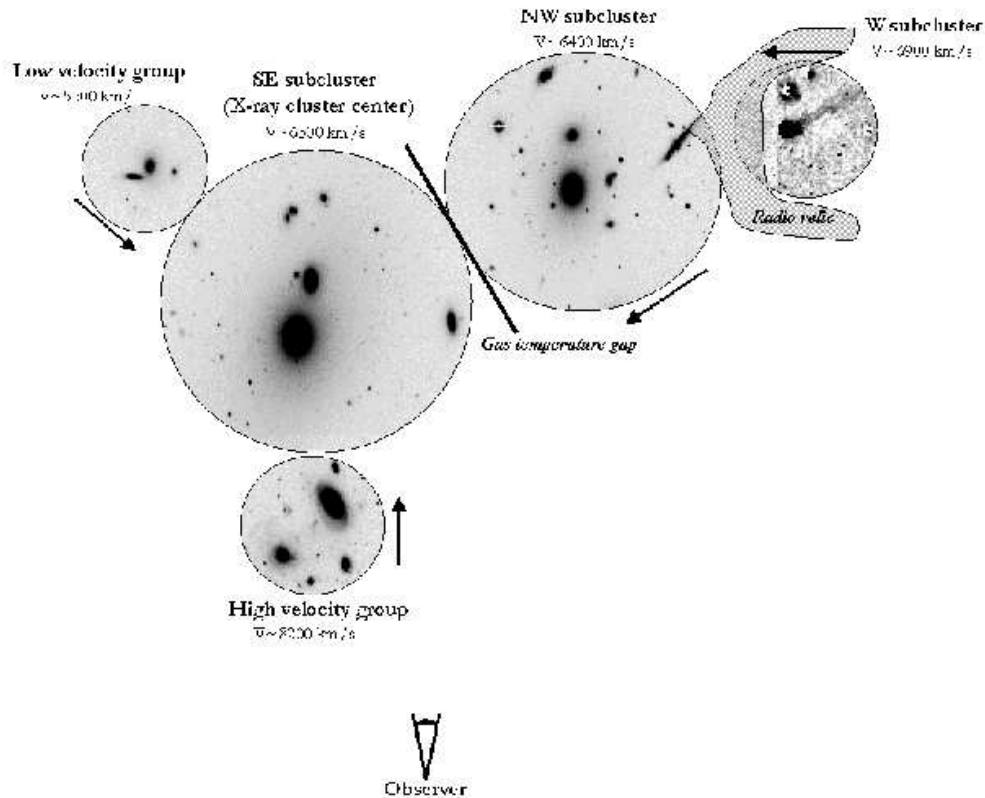}
\caption{A 3D sketch of Abell 1367 summarizing the various sub-components described in Section \ref{galdistr}.
The cluster is viewed from its near side, as suggested by the eyeball indicating the observer's position.}
\label{cartoon}
\end{figure*}
\begin{figure}[t]
\centering
\includegraphics[width=8.5cm]{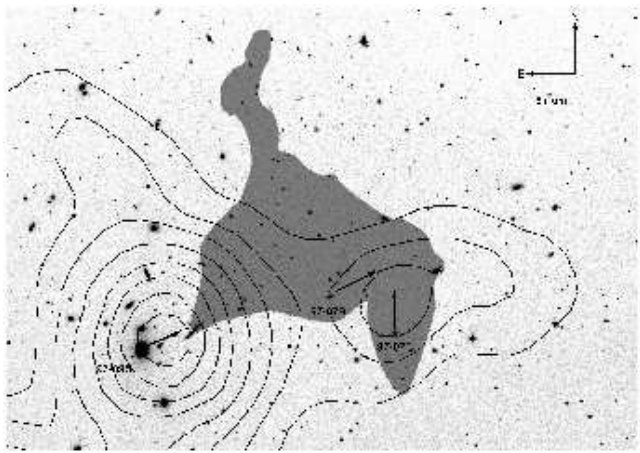}
\caption{Blow-up of the NW substructure of Abell 1367. The arrows indicate the direction of 
radio head tails associated with 97-079 and 97-073 and the orientation of the NAT radio galaxy 
97-095. The dashed region shows the distribution of the diffuse cluster radio relic \citep{GAV78}.
The iso-density contours for the 146 confirmed cluster members are superposed.}
\label{NWsub}
\end{figure}
\subsection{The North-West subcluster}
\begin{figure}[t]
\centering
\includegraphics[width=8.5cm]{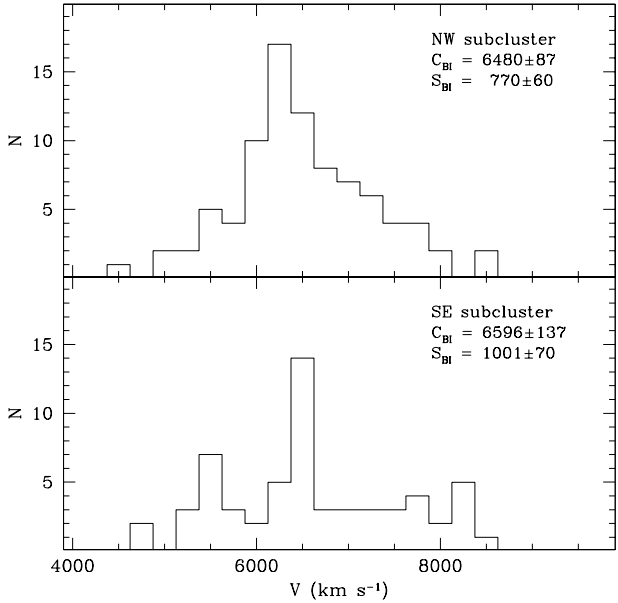}
\caption{The LOS velocity distribution for galaxies in the NW (upper) and in the SE 
(lower) subclusters.}
\label{sub_histvel}
\end{figure}

\begin{figure}[t]
\centering
\includegraphics[width=8.5cm]{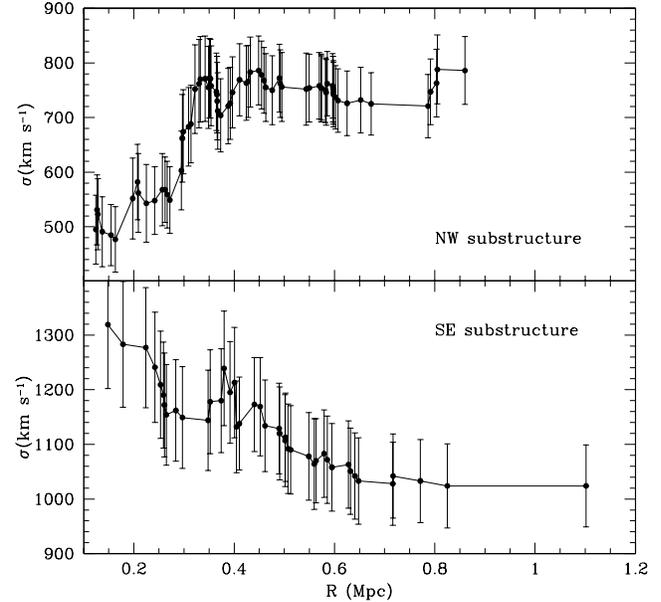}
\caption{The velocity dispersion radial profile of the NW (upper) and the SE 
(lower) subclusters.}
\label{velprof}
\end{figure}

The NW subcluster is composed of 86 galaxies and includes two density peaks: 
the highest and a 
secondary one located at the western periphery of the subcluster (labeled as W subcluster 
in Fig.\ref{cartoon}), with a weak X-ray counterpart.
It has a similar mean location ($C_{BI} = 6480\pm87 ~\rm km~s^{-1}$) and a lower 
scale ($S_{BI} = 770\pm60~\rm km~s^{-1}$) than the whole cluster.\\
Fig.\ref{sub_histvel} shows the LOS velocity distribution of this subcluster.
The W test rejects the Gaussian hypothesis at a confidence level of 39\%.
Thus the LOS velocity distribution is consistent with a 
Gaussian distribution, suggesting that this subcluster is a virialized system. 
Moreover its increasing velocity dispersion profile (see Fig. \ref{velprof}) is consistent 
with a relaxed cluster undergoing two body relaxation in the dense central region, 
with circular velocities in the center and more isotropic velocities in the external 
regions \citep{GIRARD98}. \\
However this subcluster also shows some evidences of merging (see Fig.\ref{NWsub}).
The brightest galaxy of this cloud CGCG97-095 (NGC3842), located $\sim$2 arcmin SE from 
the NW density peak, is a radio galaxy classified as a narrow-angle 
tail (NAT) \citep{BLIRIZ98}. 
The tail orientation (indicated with an arrow in  Fig. \ref{NWsub}) suggests that this galaxy (and 
the associated substructure) is moving 
from north-west to south-east, toward the main cluster core.\\
Moreover two CGCG \citep{ZWHE61} galaxies, 97-073 and 97-079, 
show features consistent with the infall scenario.
\cite{GAVCO95,GAVB01} found that both galaxies 
have their present star formation enhanced along peripheral 
HII regions which developed at the side facing the direction of motion 
through the cluster IGM. 
Their neutral hydrogen is significantly displaced in the opposite side 
\citep{DIGAV91}, 
where 50 kpc long tails are detected both in the light of the synchrotron radiation 
\citep{GAVJAF87} 
and in H$\alpha$ \citep{GAVB01}. 
The observational scenario is consistent with the idea that ram-pressure 
\citep{GUNG72} is enhancing for a limited amount of time 
the star formation of galaxies that are entering the 
cluster medium for the first time.\\ 
However these two galaxies appear not directly 
associated with the center of the NW subcluster since they lie at a projected 
distance of $\sim$0.34 Mpc from the main density peak (see Fig.\ref{NWsub}). 
Moreover their large distance ($\sim$0.48 Mpc) from 
the shock front observed in X-ray between the 
NW and the SE substructure indicates that these objects do not belong to 
the main galaxy density peak infalling into the cluster center.
Conversely they are at a projected distance of only 0.08 Mpc from the center 
of the W subcluster, suggesting 
that they are associated with this subcloud.\\
For these reasons we consider 
an alternative scenario in which these two galaxies belong to a secondary 
substructure infalling into the NW substructure from the western side (see Fig. \ref{cartoon}).
This picture is supported by the presence of the extended radio relic detected both in 
X-ray and radio continuum in this region \citep{GAV78,GAVTRI83}. 
Cluster radio halos contain fossil radio plasma, the former outflow of a radio galaxy, that has been revived 
by shock compression during cluster merging \citep{ENS98,ENS02}. 
The radio relic observed in Abell 1367 
extends, south-west to north-east, from 97-073  to 127-040 with 
a projected extent of 0.8 Mpc (see Fig.\ref{NWsub}). The age of its electrons is estimated
to be $\sim 0.2~ \rm Gyr$ \citep{ENS98}.
The only plausible source of high energy electrons available in this region is the NAT galaxy 
97-095, presently at $\sim$0.25 Mpc from the relic and whose
tails point exactly in the relic direction.  
Assuming that the fossil radio halo originated from 97-095, 
we find that the infall velocity of this galaxy into the SE subcluster is
$V \sim 1250 ~\rm km~s^{-1}$, consistent with the typical infall velocity of cluster galaxies. 
Thus the presence of the radio relic results consistent with a merging scenario in which 
the W subcluster, containing 97-079 and 97-073, is infalling into the NW substructure, 
compressing the plasma ejected from 97-095 and re-accelerating the electrons to relativistic energies.

\subsection{The South-East subcluster}
\label{SE}
The SE cloud is composed of 60 galaxies associated with the X-ray cluster center.
It has the highest LOS velocity and dispersion of the whole sample 
(see Fig.\ref{velfield}) with a location $C_{BI} = 6596\pm137 \rm~km~s^{-1}$ and a 
scale $S_{BI} = 1001\pm70~\rm km~s^{-1}$. 
Its velocity distribution, shown in Fig. \ref{sub_histvel}, appears significantly non-Gaussian.
The W test rejects the Gaussian hypothesis at a confidence level of 96.8\%,
supporting the idea that the cluster center is far 
from relaxation.
This is in agreement with the decreasing velocity dispersion profile of this 
region (see Fig.\ref{velprof}), 
consistent with isotropic velocities in the center and radial velocities 
in the external regions, as expected in the case 
of galaxy infall onto the cluster \citep{GIRARD98}.\\
The velocity distribution of Fig. \ref{sub_histvel} has three 
peaks at $\sim5500~\rm km~s^{-1}$, $\sim6500~\rm km~s^{-1}$ and $\sim8200~\rm km~s^{-1}$ respectively, 
probably associated with three separate groups. 
Moreover we remark that the galaxy gaps between the three peaks 
are fairly consistent with two of the most significant weighted gaps detected in the global 
velocity distribution ($V\sim5800~\rm km~s^{-1}$ and $V\sim7500~\rm km~s^{-1}$).\\
In order to check for any position-velocity segregation, 
we divide the SE subcluster in three groups according to their LOS velocity: 
galaxies with $V<5800~\rm km~s^{-1}$ belong to the low velocity 
group, galaxies with $V>7500~\rm km~s^{-1}$ belong to the high velocity group and 
galaxies with intermediate velocity belong to the SE subcluster.\\
\begin{figure}[t]
\centering
\includegraphics[width=9.0cm]{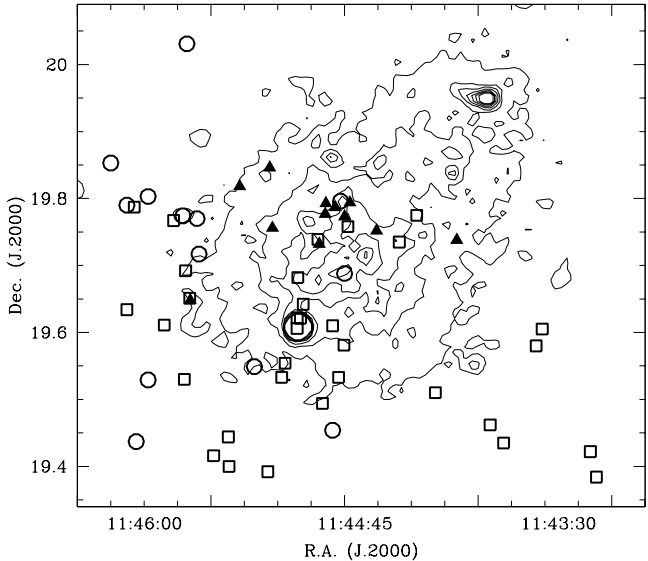}
\caption{The distribution of galaxies belonging to the South-East subcluster. 
Triangles indicate galaxies with LOS velocity $> 7500 \rm~km~s^{-1}$, circles 
galaxies with LOS velocity $< 5800 \rm~km~s^{-1}$ and squares  
galaxies with LOS velocity comprises in the range $5800\rm ~km~s^{-1} <V<7500 \rm~km~s^{-1}$. 
The ROSAT X-ray contours are shown.}
\label{SE_sky}
\end{figure}
The projected distribution of the three groups is shown in Fig.\ref{SE_sky}. 
The high-velocity group ($V\sim8200~\rm km~s^{-1}$, triangles) appears segregated 
in the northern part of the SE cloud, extending $\sim$20 arcmin in right ascension but 
only $\sim$7 arcmin in declination.
It is associated with the substructure detected by the $\Delta$ test 
(see Fig. \ref{dressler})
and with the infalling group of star-forming galaxies recently discovered 
by \cite{SAKK02} and by \cite{GAVC03}.
Its spatial segregation and high star formation activity suggest that this group is a 
separate unit infalling into the cluster, probably from the near side (see Fig. \ref{cartoon}).
It is remarkable that \cite{SUNM02}, using Chandra observations of the cluster center, discovered 
a ridge-like structure around the cluster center, $\sim$6 arcmin south from the center of the 
high velocity group, probably associated with a compact merging subcluster 
(perhaps this group) penetrating the SE cluster core.\\
The low-velocity group ($V\sim5500~\rm km~s^{-1}$, circles in Fig.\ref{SE_sky}) 
seems segregated in the eastern part of the cloud, perhaps infalling 
from the eastern side into the cluster core (Fig. \ref{cartoon}). 
This scenario is also supported by the detection of
cool gas streaming into the cluster core from the eastern side (\citealp{FORCH03}), 
probably associated with this low velocity group of galaxies.\\
Galaxies with $V\sim6500~\rm km~s^{-1}$ (squares in Fig.\ref{SE_sky}) 
are homogeneously distributed over the SE subcluster, representing 
its virialized galaxy population.
However the brightest galaxy in this group 97-127 (NGC3862) is a NAT radio galaxy with very 
extended radio tails pointing in the direction of the low velocity group \citep{GAVPER81}, 
suggesting motion relative to the IGM.\\ 
The velocity-space segregation observed in the SE subcluster 
suggests that the cluster center is experiencing multiple merging of at least 
two separate groups, supporting the idea that it is far 
from relaxation. This picture is consistent with the high gas entropy in this region, 
since in absence of a cool dense core the substructures infalling into 
the major cluster can penetrate deep inside, disturbing the cluster 
core dynamics \citep{CHUFO03}.\\
A sketch of the various substructures identified in Abell 1367 by the present study, is given in Fig. \ref{cartoon}.
Five substructures are detected. Two clouds, the NW and SE subclusters, are 
in the early merging phase, meanwhile three smaller groups are infalling into Abell 1367. 
The W subcloud, associated with 
the head-tail systems 97-073/79, is probably infalling into the NW subcluster, exciting  
the radio relic observed in between the two structures. 
The other two groups are infalling into the SE subcluster: the low velocity group from the eastern side, 
while the high velocity group from the near side.
\begin{figure}[t]
\centering
\includegraphics[width=8.5cm]{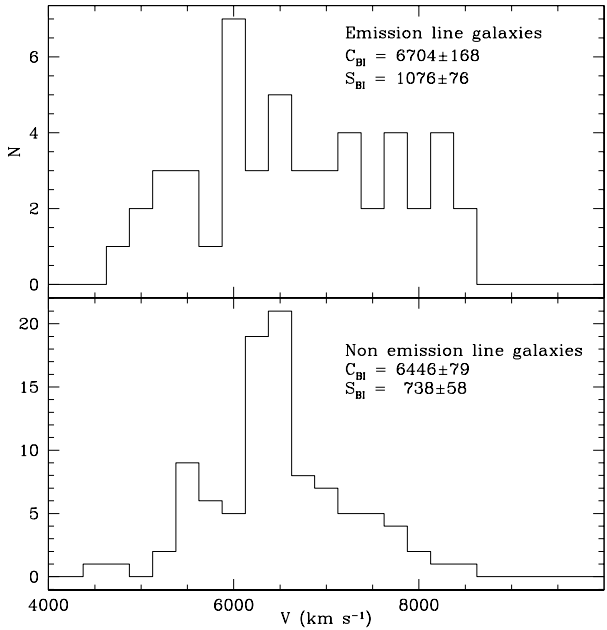}
\caption{The LOS velocity distribution for emission line (upper) and non 
emission line galaxies (lower) in the whole cluster sample.}
\label{typehist}
\end{figure}

\section{Star formation activity in the infalling groups}
\label{star}
\begin{figure*}[t]
\centering
\includegraphics[width=15.0cm]{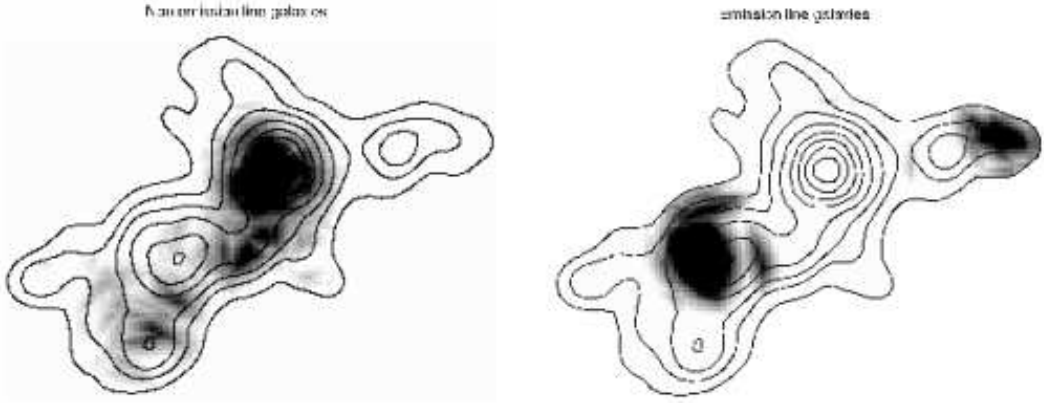}
\caption{Projected density map of non emission line (left) and emission line (right) 
galaxies in Abell 1367. The iso-density contours of the 146 confirmed 
cluster members are superposed.}
\label{skytype}
\end{figure*}

The dynamical study presented in the previous sections indicates that 
Abell 1367 is a dynamically young cluster in the early stage of a multiple merging event
involving at least five substructures.
Since merging is expected to trigger star formation in cluster galaxies 
\citep{BEKKI99}, we study separately the spatial and velocity distribution of the star forming galaxies. 
Only 49 out of the 146 cluster members show recent star formation activity (e.g. H$\alpha$ line in 
emission, \citealp{IPB02,GAVB03a}; Cortese et al., in preparation). 
Fig.\ref{typehist} shows the LOS velocity distribution of galaxies divided into emission line 
(upper panel) and non emission line (lower panel) galaxies.
The star forming sample has higher location and scale ($C_{BI} = 6704\pm168\rm~km~s^{-1}$, 
$S_{BI} = 1076\pm76\rm~km~s^{-1}$) than the quiescent sample 
($C_{BI} = 6446\pm79\rm~km~s^{-1}$, $S_{BI} = 738\pm58\rm~km~s^{-1}$).  
According to a two-sample Kolmogorov-Smirnov test the two velocity distributions 
have only $\sim$5\% probability of being consistent, suggesting a different origin and/or evolution.
We remark that, if the star forming galaxies are infalling onto the cluster along 
radial orbits, their velocity dispersion should be $\sim \sqrt{2}$ times the velocity dispersion 
of the relaxed sample, as observed in this case. 
This result suggests that star forming systems are an infalling population 
while the non-star forming galaxies represent the virialized cluster population.\\
The projected density distribution of star forming and non star forming 
is shown in Fig.\ref{skytype}. The highest density of non emission line systems is 
observed near the center of the NW substructure.
This morphological segregation further supports the idea that the NW cloud is a relaxed system 
merging for the first time into the SE subcluster.\\
The emission line galaxies have a different distribution.
The highest density of star forming systems is in the infalling groups, i.e. in the high velocity 
group infalling into the SE subcluster and in the W cloud infalling into the NW 
substructure, suggesting that their interaction with the cluster environment 
is triggering some star formation activity.
Indeed in these systems the fraction of star forming galaxies lies 
between 64\% and 36\%, decreasing to 31\% in the NW substructure and to 20\% in the SE subcluster.
  
\section{Cluster mass}
\label{mass}
\begin{table}  
\caption{Mass estimate for Abell 1367} 
\label{Mass} \[ 
\begin{tabular}{lccc} 
\hline 
\noalign{\smallskip}   
Sample & $\rm R_{H}$ & $\rm M_{V}$ & $\rm M_{PM}$ \\      
	 & Mpc & $\rm 10^{14} M_{\odot}$ & $\rm 10^{14} M_{\odot}$ \\ 
\noalign{\smallskip} 
\hline 
\noalign{\smallskip} 
 A1367~all~types         & 0.41 &  7.04 $\pm$ 0.90  & 7.82 $\pm$ 2.50 \\
 A1367~non-star~forming  & 0.37 &  4.35 $\pm$ 0.70  & 5.11 $\pm$ 0.90 \\
 A1367~NW~all~types      & 0.30 &  3.87 $\pm$ 0.62  & 6.12 $\pm$ 1.52 \\
 A1367~NW~non-star~forming     & 0.24 &  2.47 $\pm$ 0.46  & 3.29 $\pm$ 0.59 \\
 A1367~SE~all~types            & 0.27 &  5.80 $\pm$ 0.88  & 6.87 $\pm$ 1.20 \\
 A1367~SE~non-star~forming     & 0.26 &  3.90 $\pm$ 0.83  & 5.58 $\pm$ 0.74 \\
\noalign{\smallskip} 
\hline 
\end{tabular} \] 
\end{table}  
The virial theorem is the standard tool 
used to estimate the dynamical mass of galaxy clusters. 
Under the assumptions of spherical symmetry and hydrostatic equilibrium and if the mass 
distribution follows the distribution of the observed galaxies independent 
of their luminosity, the total gravitational mass of a cluster is given by
\begin{displaymath}
M_{V} = \frac{3\pi}{G}\sigma^{2} R_{H}
\end{displaymath}
where $\sigma$ is the galaxy velocity dispersion and 
$R_{H}$ is the cluster mean harmonic radius:
\begin{displaymath}
R_{H} = \frac{N(N-1)}{\sum_{i>j} R_{ij}^{-1}}
\end{displaymath}
where $N$ is the total number of galaxies.\\
An alternative approach is to use the projected mass estimator \citep{HETRE85}, 
defined as
\begin{displaymath}
M_{PM} = \frac{32}{\pi G N} \sum_{i} V_{i}^{2} R_{i}
\end{displaymath}
where $V_{i}$ is the observed radial component of the velocity of the $i$ galaxy 
with respect to the systemic cluster velocity, and $R_{i}$ is its projected separation
from the cluster center.
The numerical factor 32 assumes that galaxy orbits are isotropic. 
In case of purely radial or purely 
circular orbits this factor becomes 64 or 16 respectively.\\
Mass estimates obtained using the two above methods and their uncertainties are 
listed in Table \ref{Mass}.
We remark that these mass estimates are probably biased 
by the dynamical state of Abell 1367, which appears far from virialization.
In particular the presence of substructures leads to an overestimate of the 
cluster mean harmonic radius and velocity dispersion, and thus of the 
virial mass (\citealp{PINK96}).
For this reason the mass derived for the whole cluster and 
for the SE and NW subclusters separately is probably overestimated.
Assuming that the early type sample represents the virialized cluster population 
(see previous section), we also derive mass estimates for the three dynamical units using 
the non star forming systems only.\\
For all the studied samples the viral mass estimates are affected 
by smaller uncertainties and yield smaller values than the projected
mass estimates. 
This can be due to the contamination by interlopers 
\citep{HETRE85} or, more probably, to the assumption of isotropic orbits.
Indeed assuming purely radial or circular orbits the mass estimate varies by a factor of 2, becoming consistent
with the virial mass.\\ 
The mass inferred from the non-star forming population are, as expected, systematically 
lower than the ones obtained from all types.
The value obtained for the whole sample is consistent with the mass estimates 
available in the literature ($M_{V}=7.26\pm1.40~ 10^{14} \rm M_{\odot}$ \citealp{GIRARD98}; 
$M_{V}=6.07\pm0.93 ~10^{14} \rm M_{\odot}$, $M_{PM}=6.28\pm0.80 ~10^{14} \rm M_{\odot}$ 
\citealp{RINGE03}).

\section{Two-Body Analysis}
\label{2b}
\begin{figure*}[t]
\centering
\includegraphics[height=20.15cm]{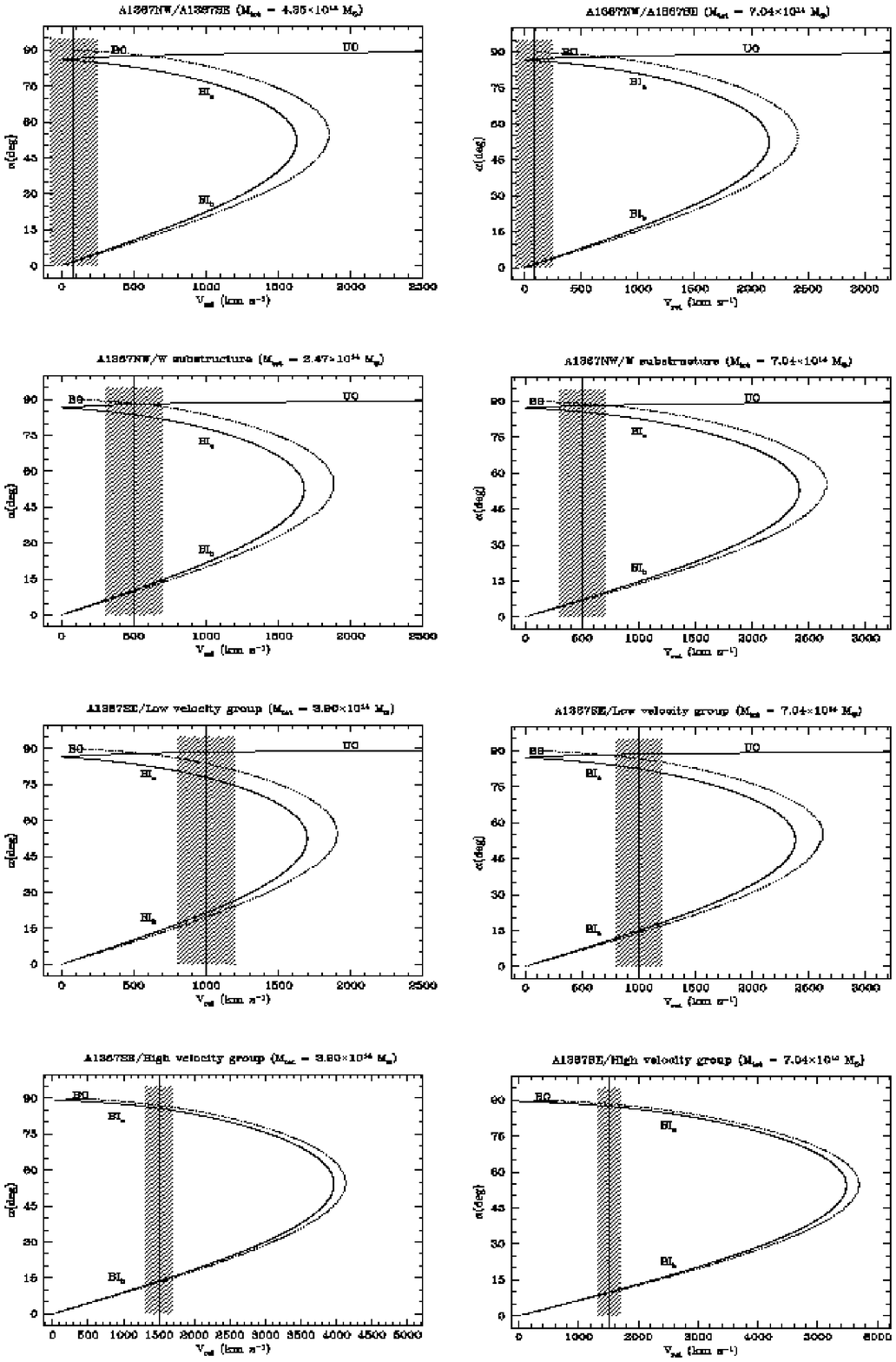}
\caption{The bound and unbound orbit regions in the ($V_{rel}, \alpha$) plane. 
The bound-incoming solutions ($BI_{a}$ and $BI_{b}$), the 
bound-outgoing solutions ($BO$) and the unbound-outgoing (UO) solutions are 
indicated with solid lines.
The dotted lines show the dividing line between bound and unbound regions. 
The vertical solid lines represent the observed $V_{rel}$ and the dashed 
regions their associated $1\sigma$ uncertainty. }
\label{2body}
\end{figure*}

In this section we investigate whether the two clouds A1367NW,  
A1367SE and the three groups infalling into the SE and NW subclusters 
form gravitationally bound systems.
For each system we apply the two-body analysis described by \cite{BEGE91}.
The two subclumps are treated as point masses moving on radial orbits.
They are assumed to start their evolution at time t=0 with zero 
separation, and are moving apart or coming together for the first time 
in their history. For bound radial orbits, the parametric 
solutions to the equations of motion are:
\begin{displaymath}
R=\frac{R_{m}}{2}(1-\cos\chi)
\end{displaymath} 
\begin{displaymath}
t=\Big(\frac{R_{m}^{3}}{8GM}\Big)^{1/2}(\chi - \sin\chi)
\end{displaymath} 
\begin{displaymath}
V=\Big(\frac{2GM}{R_{m}}\Big)^{1/2} \frac{\sin\chi}{(1-\cos\chi)}
\end{displaymath} 
where $R$ is the components separation at time $t$, and $V$ is 
their relative velocity. $R_{m}$ is the separation of the subclusters at maximum 
expansion and $M$ is the total mass of the system.
Similarly, the parametric solutions for the unbound case are:
\begin{displaymath}
R=\frac{GM}{V_{\infty}^{2}}(\cosh\chi - 1)
\end{displaymath} 
\begin{displaymath}
t=\frac{GM}{V_{\infty}^{3}}(\sinh\chi - \chi)
\end{displaymath} 
\begin{displaymath}
V=V_{\infty} \frac{\sinh\chi}{(\cosh\chi - 1)}
\end{displaymath} 
where $V_{\infty}$ is the asymptotic expansion velocity.\\
The system parameters $V$ and $R$ are related to the observables 
$V_{rel}$ (the LOS relative velocity) and $R_{p}$ (the projected separation)
by:
\begin{displaymath}
V_{rel}= V \sin \alpha,	~~~~~	R_{p}= R \cos \alpha
\end{displaymath} 
where $\alpha$ 
is the angle between the plane of the sky and the line joining the centers of the two 
components.
The two systems are thus closed 
by setting the present time 
to $t_{0}=13$ Gyr (the age of the Universe in a $\Omega_{m}$=0.3 and $\Omega_{\lambda}$=0.7 cosmology)
and solved iteratively to determine the projection angle 
as a function of $V_{rel}$.\\
We determine two solutions for each two-body model, assuming two extreme values 
for the total mass of each system ranging from  the virial mass 
of the non-star forming population to the virial mass of the whole cluster.
Table \ref{twobodypar} summarizes the adopted parameters of the two-body analysis, and 
Fig. \ref{2body} shows the computed solutions in the ($\alpha,V_{rel}$) plane.
The vertical lines represent the observed values of $V_{rel}$ and 
the dashed regions their associated $1\sigma$ uncertainties.\\
The solutions have three different regimes:
an unbound-outgoing regime (UO), a bound-outgoing regime (BO) and a bound-ingoing regime (BI).
It is easy to show 
that the unbound solutions will lie in the region of the ($\alpha,V_{rel}$) plane where:
\begin{displaymath}
V_{rel}^{2} R_{p} \leq 2GM_{tot} \sin^{2} \alpha \cos \alpha.
\end{displaymath} 
The dotted lines in Fig. \ref{2body} show the dividing line between bound and unbound 
regions.\\
In the BO regime, the two subclumps are still separating and have not yet reached 
the maximum expansion.\\
The BI regime describes the system after maximum 
expansion. 
For each $V_{rel}$, there are two corresponding values of $\alpha$, a large 
and a  small one. The large value assumes that the substructures are far apart, 
with low relative velocity, while the small value implies that the subclusters are close together 
near the plane of the sky (see Fig. 7 in \citealp{BEGE91}).
Thus we split the BI regime into two branches, called $BI_{a}$ and $BI_{b}$.\\
The probability of each solution, computed following the procedure described by 
\cite{BEGE91}, is given in Table \ref{twobodypar}. 
Our result is that the \emph{A1367NW/SE} and the \emph{A1367SE/High Velocity group} 
systems are bound with 100\% probability and presently infalling 
with 96\% and 100\% probability respectively.
The \emph{A1367NW/W} and the \emph{A1367SE/Low Velocity group} systems are bound at 99\% 
and 96\% probability respectively.         
We conclude that all systems constituting Abell 1367 are gravitationally bound at $\geq$ 96\% probability.

\begin{table*}  
\caption{Two-body model parameters} 
\label{twobodypar} \[ 
\begin{tabular}{lccccccc} 
\hline 
\noalign{\smallskip}   
System & $\rm M_{tot}$ & $\rm V_{rel} \pm \Delta V_{rel}$ & $\rm R_{p}$  & \multicolumn{4}{c}{Solution Probability} \\      
       &  &  &  & $\rm BI_{a}$ &  $\rm BI_{b}$ & BO & UO   \\ 
       & $\rm 10^{14} ~M_{\odot}$ & $\rm km~ s^{-1}$ & Mpc & \% &  \% & \% & \%   \\ 
\noalign{\smallskip} 
\hline 
\noalign{\smallskip} 
A1367NW/SE   	      	  &   7.04  &  $84 \pm 162$ & 0.45 & 57 & 40 & ~3 & ~0 \\
                      	  &   4.35  &  $84 \pm 162$ & 0.45 & 55 & 41 & ~4 & ~0 \\
A1367NW/W    	      	  &   7.04  &  $500\pm 200$ & 0.37 & 57 & 40 & ~2 & ~1 \\
             	      	  &   2.47  &  $500\pm 200$ & 0.37 & 56 & 41 & ~2 & ~1 \\
A1367SE/Low~ Vel.~ gr.    &   7.04  & $1000\pm 200$ & 0.38 & 58 & 40 & ~0 & ~2 \\
             	          &   3.90  & $1000\pm 200$ & 0.38 & 57 & 39 & ~0 & ~4 \\
A1367SE/High~ Vel.~ gr.   &   7.04  & $1500\pm 200$ & 0.08 & 56 & 44 & ~0 & ~0 \\
             	          &   3.90  & $1500\pm 200$ & 0.08 & 58 & 42 & ~0 & ~0 \\
\noalign{\smallskip} 
\hline 
\end{tabular} \] 
\end{table*}  

\section{Conclusions}
\label{concl}
We present a dynamical analysis of the central $\sim$ 1.3 square degrees of the 
galaxy cluster Abell 1367, based on 273 redshift of which 119 are new measurements.
The LOS velocity distribution of the 146 cluster members 
is significantly non Gaussian, suggesting 
that the cluster is dynamically young.
The member galaxies show an elongated distribution along 
the NW-SE direction with two major density peaks, 
consistent with the X-ray morphology.
The strong difference in the LOS velocity and velocity dispersion 
of the two density peaks, the abrupt gas temperature 
gradient detected in X-rays and the 3D statistical tests  
support a merging scenario involving at least two subclusters.
Moreover the dynamical properties of the NW and SE clouds suggest 
an even more complex picture, summarized in Fig. \ref{cartoon}.
At least another group of star forming galaxies (the high velocity group) infalling into the cluster core is 
detected, suggesting a multiple merging event.
Furthermore our analysis suggests the presence of two
other groups infalling into the cluster center.
In the North-West part of Abell 1367 a group of galaxies (W subcluster), 
associated with the infalling galaxies 97-073/79 
and with the radio relic observed in this region, 
is probably merging with the relaxed core of the NW subcluster.
In the South part another group (the low velocity group) is infalling from the eastern side 
into the disturbed core of the SE subcluster.
These three subgroups have a higher fraction of star forming galaxies
than the cluster core, as expected during the 
early phase of merging events.\\
The multiple merging scenario is consistent with the location of Abell 1367
being at the intersection of two filaments, the first extending 
roughly 100 Mpc from Abell 1367 toward Virgo \citep{WEST00} and the second extending between 
Abell 1367 and Coma (as a part of the Great Wall, \citealp{ZAGE93}).
As predicted by \cite{KATWH93} this is the natural place for Abell 1367 
to evolve into a rich relaxed cluster.

\begin{acknowledgements}
We thank the referee, R. Hank Donnelly, for his useful comments 
which helped us to improve and strengthen the paper.      
We thank Paolo Franzetti for his help during the preparation of 
the AF2-WYFFOS configurations and Timothy Beers for kindly providing 
the program ROSTAT. 
The TACS of the William Herschel, Loiano and Cananea telescopes 
are acknowledged for the generous time allocation to this project. 
This work could not be completed without the NASA/IPAC Extragalactic Database (NED) which 
is operated by the Jet Propulsion Laboratory, Caltech under contract with NASA.
This research has made use of the GOLD Mine Database, operated by the Universit\'{a} degli Studi di Milano-Bicocca. 
\end{acknowledgements}

\onecolumn
\begin{table*}  
\caption{The 119 new redshift measurements} 
\label{redshift} \[ 
\scriptsize
\begin{tabular}{ccccccccccccc} 
\hline 
\noalign{\smallskip}   
\hline 
\noalign{\smallskip}   
    Name & R.A.      & Dec.      & r'  &  V    &  Tel. & ~~  &  Name & R.A.	 & Dec.      & r'  &  V    &  Tel.\\	  
          & (J.2000) & (J.2000) & mag &  $\rm km \ s^{-1}$ &  & ~~  &       & (J.2000) & (J.2000) & mag &  $\rm km \ s^{-1}$ &		\\  
\noalign{\smallskip} 
\hline 
\noalign{\smallskip}
\hline 
114000+195426 &   114000.62  &	  195426.7   &	 15.98 & 10883 & CAN & ~~  & 114356+201404 &   114356.80  &    201404.9   &   18.42 & 72058 & WHT \\
114159+193227 &   114159.52  &	  193227.3   &	 15.63 & 21228 & CAN & ~~  & 114357+201122 &   114357.69  &    201122.7   &   17.06 & 5348  & WHT \\
114200+195846 &   114200.83  &	  195846.0   &	 17.09 & 6420  & WHT & ~~  & 114358+195330 &   114358.86  &    195330.2   &   19.22 & 6200  & WHT \\
114208+191905 &   114208.01  &	  191905.0   &	 19.04 & 23456 & WHT & ~~  & 114359+195630 &   114359.51  &    195630.8   &   20.37 & 6992  & WHT \\
114212+195650 &   114212.47  &	  195650.3   &	 17.73 & 20278 & WHT & ~~  & 114402+194742 &   114402.65  &    194742.7   &   17.52 & 43665 & WHT \\
114213+193001 &   114213.87  &	  193001.6   &	 16.92 & 23641 & WHT & ~~  & 114403+200552 &   114403.70  &    200552.6   &   15.80 & 5698  & WHT \\
114215+200427 &   114215.59  &	  200427.0   &	 19.20 & 6100  & WHT & ~~  & 114404+192922 &   114404.17  &    192922.8   &   18.59 & 53335 & WHT \\
114219+200548 &   114219.15  &	  200548.0   &	 16.45 & 6841  & CAN & ~~  & 114404+195956 &   114404.65  &    195956.6   &   17.33 & 33830 & WHT \\
114224+195329 &   114224.39  &	  195329.8   &	 18.29 & 31440 & WHT & ~~  & 114407+193850 &   114407.21  &    193850.9   &   17.10 & 20877 & WHT \\
114224+191157 &   114224.48  &	  191157.0   &	 16.39 & 28546 & CAN & ~~  & 114407+193143 &   114407.71  &    193143.1   &   18.44 & 53424 & WHT \\
114226+194317 &   114226.24  &	  194317.1   &	 17.50 & 23416 & WHT & ~~  & 114412+195503 &   114412.22  &    195503.9   &   17.65 & 20916 & WHT \\
114230+191447 &   114230.62  &	  191447.5   &	 18.60 & 27304 & WHT & ~~  & 114412+195633 &   114412.27  &    195633.4   &   17.02 & 6244  & WHT \\
114230+192553 &   114230.95  &	  192553.8   &	 17.80 & 45683 & WHT & ~~  & 114412+201119 &   114412.92  &    201119.7   &   19.25 & 74731 & WHT \\
114238+194718 &   114238.24  &	  194718.6   &	 17.04 & 25610 & WHT & ~~  & 114415+193037 &   114415.25  &    193037.5   &   16.58 & 6502  & WHT \\
114239+195145 &   114239.78  &	  195145.9   &	 19.06 & 53710 & WHT & ~~  & 114415+193012 &   114415.33  &    193012.3   &   18.27 & 35227 & WHT \\
114240+195627 &   114240.26  &	  195627.5   &	 18.63 & 19946 & WHT & ~~  & 114417+194543 &   114417.28  &    194543.9   &   18.14 & 66264 & WHT \\
114243+191615 &   114243.81  &	  191615.8   &	 18.72 & 5312  & WHT & ~~  & 114422+194628 &   114422.16  &    194628.2   &   15.70 & 6527  & CAN \\
114249+193935 &   114249.85  &	  193935.1   &	 19.14 & 72429 & WHT & ~~  & 114426+195951 &   114426.10  &    195951.5   &   16.98 & 30102 & WHT \\
114250+193955 &   114250.47  &	  193955.7   &	 19.22 & 13759 & WHT & ~~  & 114430+194258 &   114430.30  &    194258.3   &   18.78 & 40347 & WHT \\
114252+195656 &   114252.17  &	  195656.4   &	 16.69 & 5936  & WHT & ~~  & 114432+195341 &   114432.19  &    195341.6   &   18.89 & 42649 & WHT \\
114254+193851 &   114254.40  &	  193851.3   &	 17.17 & 6406  & WHT & ~~  & 114432+194734 &   114432.98  &    194734.6   &   18.82 & 71100 & WHT \\
114254+194033 &   114254.93  &	  194033.6   &	 18.87 & 71389 & WHT & ~~  & 114447+201248 &   114447.20  &    201248.5   &   18.17 & 6699  & WHT \\
114258+194321 &   114258.13  &	  194321.1   &	 18.98 & 6523  & WHT & ~~  & 114449+195628 &   114449.72  &    195628.9   &   16.70 & 5539  & WHT \\
114258+194053 &   114258.37  &	  194053.9   &	 19.25 & 71436 & WHT & ~~  & 114501+195504 &   114501.97  &    195504.5   &   18.79 & 45708 & WHT \\
114258+194644 &   114258.53  &	  194644.2   &	 19.00 & 88274 & WHT & ~~  & 114503+193831 &   114503.00  &    193831.2   &   16.76 & 6193  & LOI \\
114258+195612 &   114258.94  &	  195612.7   &	 18.41 & 7059  & WHT & ~~  & 114503+194743 &   114503.14  &    194743.9   &   17.91 & 23374 & WHT \\
114259+194801 &   114259.71  &	  194801.1   &	 18.92 & 71600 & WHT & ~~  & 114504+201412 &   114504.25  &    201412.2   &   18.31 & 5477  & WHT \\
114300+192515 &   114300.65  &	  192515.2   &	 18.42 & 53145 & WHT & ~~  & 114505+194057 &   114504.83  &    194056.9   &   15.67 & 6506  & LOI \\
114301+194758 &   114301.24  &	  194758.9   &	 18.67 & 72572 & WHT & ~~  & 114506+200849 &   114506.38  &    200849.9   &   19.23 & 3822  & WHT \\
114301+195313 &   114301.97  &	  195313.5   &	 18.61 & 46935 & WHT & ~~  & 114509+194845 &   114509.38  &    194845.4   &   17.49 & 19831 & WHT \\
114307+192807 &   114307.13  &	  192807.3   &	 17.37 & 32298 & WHT & ~~  & 114509+193316 &   114509.40  &    193316.2   &   15.80 & 7409  & LOI \\
114307+193029 &   114307.16  &	  193029.8   &	 17.93 & 23763 & WHT & ~~  & 114509+194526 &   114509.65  &    194526.9   &   16.90 & 19834 & WHT \\
114310+192526 &   114310.09  &	  192526.4   &	 16.62 & 19188 & WHT & ~~  & 114516+193245 &   114516.18  &    193245.1   &   16.80 & 19669 & WHT \\
114310+191519 &   114310.29  &	  191519.2   &	 17.41 & 23578 & WHT & ~~  & 114517+200120 &   114517.10  &    200120.7   &   15.32 & 14745 & LOI \\
114313+200747 &   114313.18  &	  200747.9   &	 16.40 & 5383  & CAN & ~~  & 114517+201108 &   114517.29  &    201108.8   &   18.21 & 79253 & WHT \\
114314+194821 &   114314.49  &	  194821.7   &	 19.29 & 71433 & WHT & ~~  & 114517+200110 &   114517.64  &    200110.0   &   15.46 & 14713 & LOI \\
114314+192534 &   114314.99  &	  192534.3   &	 15.76 & 23867 & CAN & ~~  & 114520+194220 &   114520.33  &    194220.3   &   20.44 & 54544 & WHT \\
114317+195525 &   114317.25  &	  195525.1   &	 18.88 & 30273 & WHT & ~~  & 114520+193259 &   114520.49  &    193259.4   &   17.48 & 4653  & WHT \\
114317+194658 &   114317.61  &	  194658.2   &	 15.69 & 6295  & CAN & ~~  & 114522+195146 &   114522.62  &    195146.5   &   21.14 & 18012 & WHT \\
114318+201523 &   114318.05  &	  201523.3   &	 17.81 & 46170 & WHT & ~~  & 114524+201239 &   114524.33  &    201239.3   &   18.73 & 44376 & WHT \\
114319+192520 &   114319.68  &	  192520.9   &	 15.99 & 6757  & WHT & ~~  & 114526+201056 &   114526.27  &    201056.8   &   16.40 & 20134 & CAN \\
114320+193637 &   114320.44  &	  193637.1   &	 19.71 & 44171 & WHT & ~~  & 114529+195658 &   114529.39  &    195658.2   &   16.29 & 24000 & WHT \\
114320+195206 &   114320.66  &	  195206.2   &	 18.15 & 52416 & WHT & ~~  & 114530+193639 &   114530.37  &    193639.4   &   17.20 & 40000 & LOI \\
114322+195704 &   114322.06  &	  195704.7   &	 16.94 & 7909  & WHT & ~~  & 114531+200217 &   114531.31  &    200217.5   &   19.78 & 45691 & WHT \\
114324+194121 &   114324.66  &	  194121.4   &	 18.33 & 35778 & WHT & ~~  & 114533+194505 &   114533.88  &    194505.9   &   18.11 & 31440 & WHT \\
114332+201326 &   114332.24  &	  201326.1   &	 16.46 & 33438 & CAN & ~~  & 114533+200028 &   114533.97  &    200028.7   &   17.56 & 35830 & WHT \\
114332+195108 &   114332.72  &	  195108.2   &	 19.07 & 14313 & WHT & ~~  & 114536+194253 &   114536.19  &    194253.7   &   18.60 & 48966 & WHT \\
114335+200005 &   114335.47  &	  200005.6   &	 16.38 & 20600 & CAN & ~~  & 114540+194302 &   114540.32  &    194302.8   &   17.74 & 5545  & WHT \\
114336+193930 &   114336.07  &	  193930.8   &	 19.24 & 44616 & WHT & ~~  & 114543+193854 &   114543.65  &    193854.9   &   16.93 & 7828  & LOI \\
114337+193835 &   114337.17  &	  193835.8   &	 17.31 & 12502 & WHT & ~~  & 114543+193905 &   114543.77  &    193905.9   &   16.30 & 7301  & LOI \\
114337+201533 &   114337.82  &	  201533.5   &	 20.19 & 11464 & WHT & ~~  & 114544+194013 &   114544.86  &    194013.3   &   17.18 & 19487 & WHT \\
114339+193446 &   114339.09  &	  193446.2   &	 16.01 & 7477  & WHT & ~~  & 114545+193151 &   114545.66  &    193151.4   &   18.57 & 6880  & LOI \\
114342+193636 &   114342.18  &	  193636.3   &	 19.26 & 71296 & WHT & ~~  & 114545+201200 &   114545.78  &    201200.3   &   19.11 & 27431 & WHT \\
114343+195607 &   114343.12  &	  195607.8   &	 18.56 & 19711 & WHT & ~~  & 114548+192708 &   114548.13  &    192708.4   &   16.72 & 30193 & WHT \\
114345+201252 &   114345.50  &	  201252.2   &	 19.27 & 20476 & WHT & ~~  & 114549+195915 &   114549.88  &    195915.3   &   15.87 & 20035 & CAN \\
114350+195702 &   114350.16  &	  195702.0   &	 17.98 & 6848  & WHT & ~~  & 114550+194824 &   114550.61  &    194824.6   &   18.85 & 41484 & WHT \\
114350+194138 &   114350.83  &    194138.0   &   19.13 & 72744 & WHT & ~~  & 114602+194754 &   114602.12  &    194754.3   &   19.62 & 73746 & WHT \\
114353+195004 &   114353.42  &    195004.6   &   19.23 & 27946 & WHT & ~~  & 114605+195151 &   114605.35  &    195151.0   &   18.86 & 46635 & WHT \\
114353+194422 &   114353.45  &    194422.2   &   15.66 & 6141  & WHT & ~~  & 114620+194518 &   114620.85  &    194518.0   &   17.58 & 45683 & WHT \\
114353+194315 &   114353.61  &    194315.8   &   17.17 & 23578 & WHT & ~~  &		   &		  &		  &	    &	    &	  \\
\noalign{\smallskip}
\hline 
\end{tabular} \] 
\end{table*}

\end{document}